\documentclass[prx,floatfix,superscriptaddress,nofootinbib,notitlepage]{revtex4-1}
\usepackage[utf8]{inputenc}

\usepackage{epsfig,graphics,graphicx}
\usepackage{dcolumn}
\usepackage{bm}
\usepackage{physics} 
\usepackage{color}
\usepackage{amsmath,amssymb,amsthm}
\usepackage{float}
\usepackage{centernot}
\usepackage{bbold}
\usepackage{amsfonts}
\usepackage{mathtools}
\usepackage{fullpage}
\usepackage{units}
\usepackage[usenames,dvipsnames]{xcolor}
\usepackage{braket}
\usepackage{indentfirst}
\usepackage{algorithm} 

\usepackage[T1]{fontenc}

\usepackage{mathpazo}
\usepackage{dsfont}

\usepackage{mathptmx}
\usepackage{latexsym}

\usepackage{qcircuit}

\usepackage[colorlinks]{hyperref}
\hypersetup{pdfpagemode=UseNone, allcolors=blue}

\def\>{\rangle}
\def\<{\langle}

\renewcommand{\rho}{\varrho}

\def\textbf#1{{\bf #1}}

\begin{document}

\title{{\sc Synthesis of Quantum Circuits with an \\ Island Genetic Algorithm}\\ }

\author{Fernando T. Miranda}
\affiliation{Programa de P\'os-Gradua\c{c}\~{a}o em Engenharia El\'etrica e Computa\c{c}\~{a}o\\
Universidade Presbiteriana Mackenzie, S\~{a}o Paulo, SP, Brazil.}

\author{Pedro Paulo Balbi}
\affiliation{Programa de P\'os-Gradua\c{c}\~{a}o em Engenharia El\'etrica e Computa\c{c}\~{a}o\\
Universidade Presbiteriana Mackenzie, S\~{a}o Paulo, SP, Brazil.}
\affiliation{Faculdade de Computa\c{c}\~{a}o e Inform\'atica\\
Universidade Presbiteriana Mackenzie, S\~{a}o Paulo, SP, Brazil.}

\author{Pedro C.S. Costa}
\affiliation{Quantum for New South Wales, Sydney, NSW 2000, AU}
\affiliation{School of Mathematical and Physical Sciences,
Macquarie University, Sydney, NSW 2109, AU}

\date{\today}
\begin{abstract}
While advances in quantum hardware occur in modest steps, simulators running on classical computers provide a valuable test bed for the construction of quantum algorithms. Given a unitary matrix that performs certain operations, obtaining the equivalent quantum circuit, even if as an approximation of the input unitary, is a non-trivial task and can be modeled as a search problem. This work presents an evolutionary search algorithm based on the island model concept, for the decomposition of unitary matrices in their equivalent circuit. Three problems are explored: the coin for the quantum walker, the Toffoli gate and the Fredkin gate. The algorithm proposed proved to be efficient in the decomposition of quantum circuits, and as a generic approach, it is limited only by the available computational power.
\end{abstract}
\maketitle

\section{Introduction}
\label{sec:introduction}
One of the promises of quantum computing is to make feasible the solution of problems that cannot be treated by a classic computer. According to \cite{loceff}, although the hardware that will drive quantum computing still is in its early stages of development, the theoretical fundamentals of quantum mechanics already makes it possible to build quantum algorithms, regardless of the physical implementation of these components. In particular, the decomposition of a unitary matrix into a sequence of quantum gates can be represented as an optimization problem. In \cite{williansGray} the authors suggested the use of genetic programming, and subsequently in \cite{yabuki} with genetic algorithms (GAs), to find the circuits that solve known problems, presenting as an experiment case the results for the quantum teleportation problem. With similar ideas in \cite{lukac1} they proposed a classic GA for the synthesis of arbitrary small quantum circuits and in \cite{spector} they explored the use of genetic programming (GP) as a framework for automatic quantum circuit synthesis. Finally, we can cite \cite{daskin} which uses the ideas of the evolutionary algorithm to evolve quantum circuits associated with molecular systems, and \cite{creevey2023gasp} which applies the GP for state preparation routines of quantum algorithms.

A common problem found in previous works is the use of controlled quantum gates of general unitaries, due to the difficulty of physically implementing such gates. Here, we discuss the implementation of a generic and easily scalable evolutionary-search algorithm, to find quantum circuits that yield an approximation for a given unitary matrix. The implementation allows for easy customization of the circuit components, which can be important when accessing circuit costs. In this paper, the following three important unitaries, in the quantum computation theory, are used: 

\begin{itemize}
\item The Toffoli gate, which has zero internal power dissipation, is  reversible in the classical computation sense~\cite{fredkin1982conservative}, and plays a major role in quantum algorithms, like in Shor's factoring algorithm~\cite{shor1999polynomial,van2005fast}; 

\item The Fredkin gate, a universal gate for classical reversible computation~\cite{fredkin1982conservative}, which can be used, for instance, to design a circuit for
error-correcting quantum computations~\cite{barenco1997stabilization}; and

\item The Hadamard coin, chosen from the quantum walks (QW) model, which largely employed in several quantum algorithms, such as the search algorithms in \cite{portugal2013quantum}, and which was proven to be a universal model of quantum computation \cite{childs2009universal}. The non-trivial circuit decomposition for the coin here is due to its implementation via Quantum Cellular Automata, using \cite{costa2018quantum}, which is a suitable model for physical realizations of abstract models of quantum computation. 
\end{itemize}

In the following sections, we present the framework for our solution. First, a quick introduction to the quantum circuit model of computation and the concept of universal gates are given in Section \ref{sec:uniSet}. Then, the chromosome representation, objective function, and the concept of unitary blocks are introduced. In the sequence, we discuss in detail the algorithm itself and its three main operators: crossover within a population, crossover between two populations, and mutation.
The computational setup is explained in Section \ref{section:params}. Finally, we discuss the solutions found and give possible pathways for future research. Appendices \ref{app:sol}, \ref{app:Qgates} and \ref{app:helpB} provide more information on the solutions and quantum gates used in the search.

\section{Quantum Circuits and Universal Quantum Gates}
\label{sec:uniSet}

The focus of the paper is on finding a unitary decomposition based on the circuit model of computation, which is given by a sequence of quantum gates that represent reversible computations. These gates are applied to the quantum analog of the bit, the qubits, which are quantum systems with two levels (i.e., have two states), conventionally described as computational basis $\ket{0}$ and $\ket{1}$. This representation preserves the analogy with the classical bits while contrasting mainly in that here the quantum system is allowed to be in the superposition of these states.

It is a well-known fact that the AND and NOT gates form a Universal set for classical computing, as they are able to approximate any Boolean function. A similar result can be obtained for the quantum circuit model, for which a set of gates is Universal if any unitary operation can be approximated, with arbitrary precision, by a circuit involving these gates \cite{nielsen_chuang_2010}. For instance, the gate set that generates the Clifford-group, $\{H,S,CNOT\}$, plus one non-Clifford gate, e.g. the $T$ gate,  provides a gate set for universal quantum computing \cite{PhysRevA.71.022316}. 

From a specific gate set the goal of this work is to find an equivalent circuit, through the evolutionary search, for a given unitary $U$. Moreover, we do not allow ancillary qubits for the unitary decomposition, which would allow the search to lead to approximations for a given $U$ with, say, lower depth, as carried out in \cite{jones2013low}.

The following sections explain the function that is being optimized, how the circuit is represented, and how the evolutionary algorithm works.

\section{Objective Function}
Given a target unitary matrix $U_t$ we search for the quantum circuit that produces this matrix with the greatest possible fidelity. To quantify this approximation, likewise done in~\cite{daskin}, we use the \textit{ trace fidelity} 
\begin{equation}
\label{eq:fobj}
f = \frac{1}{2^n}\left|\Tr(U_aU_t^{\dagger})\right|,
\end{equation}
where $U_a$ is the achieved unitary, $n$ is the number of qubits needed to describe $U_a$, and $\Tr(\cdot)$ is the trace operation of a given matrix. Notice that for the ideal solution, i.e., when $U_a = U_t$, $f = 1$. From the gate set used here and from implications of the \textit{Solovay–Kitaev theorem} \cite{kitaev1997quantum}, we know that very often, we can not get the ideal solution. However, allowing a circuit decomposition with deeper depth will enable us to get a more accurate approximation for a given target unitary.

It is important to note that $f$ ignores global phase differences, which is not physical, making the optimization easier. Since $U_t$ is being constructed from a gate set composed only of unitary matrices, we also have $U_t$ as a unitary matrix, entailing that $f$ is in the range $[0, 1]$. 

The circuit cost is a very important aspect of quantum computation. Instead of implementing a multi-objective optimization, which could result in an overly complex fitness function, we opted to optimize only the circuit fidelity, but we also allowed circuits with different depths to compete against each other. In this way, in the case where different circuits have the same fidelity, the circuit having the smaller depth will be considered the best solution. The maximum depth presently allowed for the optimization process is 90.

\section{Circuit Representation}
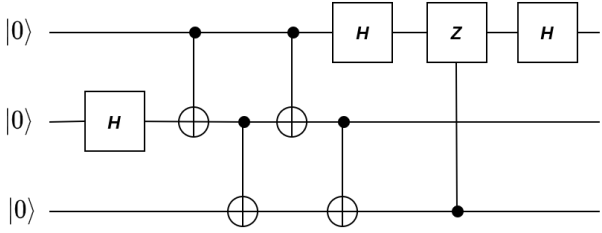
\begin{figure}[htbp]
\centerline{
\Qcircuit @R=2em @C=1em {\lstick{\ket{0}}
& \qw & \ctrl{1} & \qw & \ctrl{1} & \qw & \gate{H} & \gate{Z} & \gate {H} & \qw \\
\lstick{\ket{0}} & \gate{H} & \targ & \ctrl{1} & \targ & \ctrl{1}  & \qw & \qw & \qw &\qw\\
\lstick{\ket{0}} & \qw & \qw & \targ & \qw & \targ & \qw &\ctrl{-2} & \qw & \qw
}}
\caption{Example circuit with gates $H$, $Z$ and $CNOT$.\label{fig:circcomplex}}
\end{figure}

A string circuit representation was adopted in the work. Consider the circuit of Figure \ref{fig:circcomplex}, its chromosome 
\begin{eqnarray}
&\left(\{1: (`H', -1, None)\},\: \{1: (`X', 0, None)\},\right.\\ 
&\{2: (`X', 1, None)\}, \: \{1: (`X', 0, None)\},\nonumber\\
&\{0: (`H', -1, None), 2:(`X', 1, None)\},\nonumber\\
&\left.\{0: (`Z', 2, None)\}, \: \{0: (`H', -1, None)\}\right),\nonumber
\end{eqnarray}
where each item in this list represents (from left to right) a $`column'$ of the circuit, with each column being read as follows: {\it target qubit}: ({\it operation, controlled-qubit, optional parameter}). The absence of control is indicated by -1 and the absence of parameters by {\it None}. The optional parameter could represent, for example, the rotation angle of a generic rotation gate. Since our gate set does not include these gates, this optional parameter is always absent, being described here just for clarity. The indices are mapped from top to bottom, with 0 indicating the first qubit, 1 indicating the second, etc.

In order to speed up the search process and allow for more fine-grained control over mutation (as will be discussed in Section \ref{sec:mutat}), the concept of \textit{blocks} was created: each chromosome is made up of \textit{B} unitary blocks, each one being a circuit itself, just like described so far. This allows the injection of `helper' blocks, which may perform a specific function inside the circuit. This saves computation time because if a block has not changed, its unitary remains the same. Also, during the evolutionary search (to be detailed below) it allows for more flexibility regarding mutation operations: the set of available mutations is different for each type of block.

\begin{figure}[htbp]
\centerline{
\Qcircuit @R=1em @C=1.5em {
& \gate{H} &\ctrl{1} & \gate{H} & \qw &\ctrl{1} & \qw & \gate{H} & \ctrl{1} &  \qw \\
 & \qw & \targ & \gate{H} & \gate{S^{\dagger}} & \targ & \gate{S}  & \qw & \targ & \qw \gategroup{1}{2}{2}{2}{.7em}{--}
 \gategroup{1}{3}{2}{3}{.7em}{--}
 \gategroup{1}{4}{2}{4}{.7em}{--}
 \gategroup{1}{5}{2}{7}{.7em}{--}
 \gategroup{1}{8}{2}{9}{.7em}{--}\\
 & \mathbf{B_1}&\mathbf{B_2} & \mathbf{B_3} & &\mathbf{B_4} &  & & \mathbf{B_6}&\\  
}}
\caption{Circuit made up of five blocks.\label{fig:block}}
\end{figure}
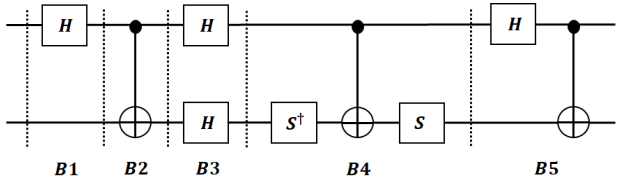

A \textit{non-controlled block} is a single column block (depth equal to one) that has no control points. A \textit{controlled block} is a single column block that has one or more control points. A \textit{helper block} is a predefined block having a depth greater than 1. Figure \ref{fig:block} illustrates the concept of blocks: B1 and B3 are non-controlled blocks, B2 is a controlled block, and B4 and B5 are helper blocks. The depth of the circuit is now given by the sum of the depths of each individual block, which may result in circuits with effectively deeper depths than when considering only the sequence of blocks.

\section{Evolutionary Algorithm}
\label{sec:EG}
When adopting a classical genetic algorithm (GA) to perform the search, the problem of premature convergence was observed: due to the lack of continuous parameters to guide the search, the exchange of genetic material tends to make individuals too similar, reaching almost always a local maximum. To overcome this problem, a GA based on the Island Model concept was adopted.

In this type of model, different populations (sets of individuals, or chromosomes) representing candidate solutions evolve independently at first. Later, they exchange genetic material in a process called {\it migration}. Thus, this model is a parallel version of the ordinary GA, where there is an exchange of information among different populations. In an intuitive way, each population may follow a different evolutionary lineage, thus increasing the genetic diversity and consequently, the effectiveness of the search \cite{whitley}.

In the first step of the algorithm, $g$ populations, each with  $p$ chromosomes, are generated in the following way: in each population, the individuals are created with a number of blocks between 4 and 15 by randomly generating each one of its blocks.
First, a block type is randomly chosen. For non-controlled and controlled blocks, gates and their positions are chosen uniformly at random. For helper blocks, the type of helper and the control wire are chosen uniformly at random.

After that, evolution will proceed in two distinct steps. The first step occurs in isolation within each population, where the best individual (the one with the highest fitness), is named the \textit{leader}, and plays the elite role during crossover. In the second step, individuals from different populations exchange genetic material in a second crossover process, ending one evolution cycle.

\begin{figure}[htb]
\begin{center}
\includegraphics[scale=0.46]{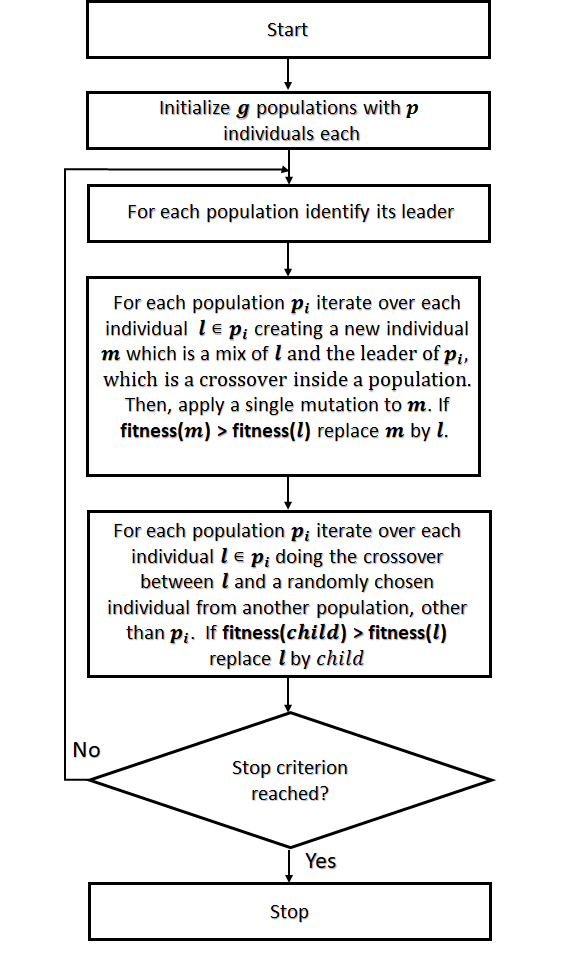}
\caption{Fluxogram of the Island model genetic algorithm.}
\label{fig:algo}
\end{center}
\end{figure}

The stop criterion employed in the search is based on the error fidelity, as also used in~\cite{PhysRevLett.93.040502}, given by 
\begin{equation}
\label{eq:error}
    \epsilon = 1 -f^2,
\end{equation}
where $f$ is defined in Eq. \eqref{eq:fobj}.
The algorithm is interrupted when Eq. \eqref{eq:error}  is smaller or equal to $1 \times 10^{-6}$. Such a stop criterion avoids an unbounded depth for each circuit decomposition. Figure \ref{fig:algo} gives an overview of the algorithm and the following sections describe the genetic operators in detail, in the order that they occur in the algorithm.

\subsection{Crossover within a population}

As depicted in Figure \ref{fig:algo}, crossover occurs at two distinct times: within a population, and in a later stage, between individuals from different populations. This section describes the crossover within the same population, which happens as follows. First, the leader individual (the one with the highest fitness) is determined. Then, for every individual of this population (which we refer to as the interacting individual), a new offspring is created by concatenating, left to right, the first $n_L$ blocks from the leader with the first $n_I$ blocks of the interacting individual. Two parameters are used to select the number of blocks involved, the ratios $r_L$ for the leader and  $r_I$  for the interacting individual, with $r_L + r_I=1$. The number of blocks extracted from the leader is given by

\begin{equation}
\label{eq:cros}
    n_L=max(1, \left \lfloor B_L \times r_L\right \rfloor),
\end{equation} 
where $B_L$ is the total number of blocks of the leader. Similarly, the number of blocks extracted from the interacting circuit is given by

\begin{equation}
\label{eq:cros2}
    n_I=max(1, \left \lfloor B_{I} \times r_I\right \rfloor),
\end{equation}
where $B_I$ is the total number of blocks of the interacting individual. In this way, it is guaranteed that the new individual will be composed of at least two blocks, and the use of $\left\lfloor \cdot \right \rfloor$ helps reduce the total depth of the offspring circuit since the \textit{floor} operation only takes into account the largest integer in the multiplication. After creation, a single mutation is applied to this new offspring. The details of the mutation operator are covered in Section~\ref{sec:mutat}. Then, if the resulting offspring has a higher fitness value, the interacting individual is replaced by this offspring; otherwise, the offspring is discarded. In this crossover, more than one child could be generated, but this would imply one more fitness check; instead, we have chosen to rely on the maximal randomness of the process, thus generating a single child.

\begin{figure}[htp]
\begin{center}
\includegraphics[scale=0.30]{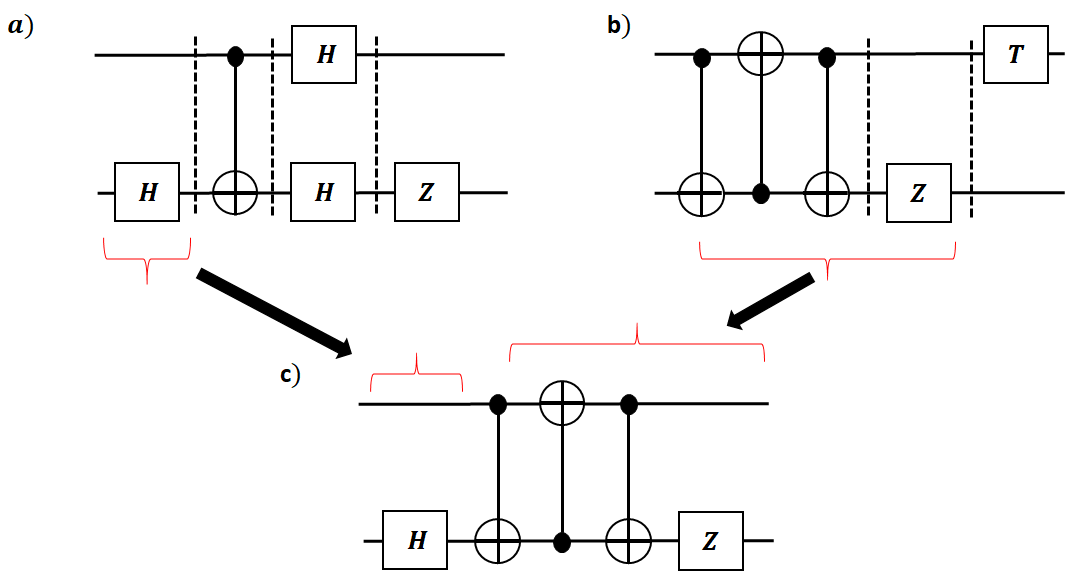}
\caption{Example of a crossover involving the leader circuit $a$ and the interacting circuit $b$. Notice that although the second and the third blocks in $b$ together just add one depth to its circuit, in our approach, they are considered two distinct blocks. The new individual $c$ is constructed out of one block coming from the leader ($H$ block) and two from the interacting individual (the $Swap$ and $Z$ blocks).}
\label{fig:cross}
\end{center}
\end{figure}

As an example of the whole process, consider the individuals from Figure \ref{fig:cross}, where the leader is on the left, with ${\it r_L}=0.3$, and interacting circuit on the right, with ${\it r_I}=0.7$. The leader circuit is made up of 4 blocks, thus supplying $max(1, \left \lfloor(4 \times 0.3\right \rfloor) = 1$ of its first blocks to the offspring. On its part, the interacting circuit, which has 3 blocks, then provides $max(1, \left \lfloor 3 \times 0.7\right \rfloor) = 2$ of its first blocks to the offspring. The resulting offspring is depicted in Figure \ref{fig:cross}-$c$.

In the next step of the algorithm, a new type of crossover takes place, this time involving different populations.

\subsection{Crossover between two populations}
In spite of being simpler than the crossover within a population, this operator has a vital role in providing genetic variability to the evolutionary process. It happens as follows. For every individual $l$ from each population $p_i$, with $i\in\{0,1\cdots,g-1\}$, we randomly select an individual $m$ from another population, say $p_j$, with $p_j\neq p_i$. 
An empty offspring is created having a total number of blocks given by

\begin{equation}
    min(B(l), B(m)),
\end{equation}
where $B(l)$ is the total number of blocks in the individual $l \in p_i$, and $B(m)$ is the total number of blocks in individual $m \in p_j$. Next, each block of this offspring will be selected from the homologous positions of individuals $l$ or $m$, with equal probability. If this offspring has a larger fitness value, it replaces $l$, otherwise, it is discarded. Again, for efficiency reasons, a single child is generated. After all populations are iterated, the stop criterion is checked and the algorithm either returns a list of solutions or continues its search as described in Figure \ref{fig:algo}.

\subsection{Mutation}
\label{sec:mutat}

The evolutionary algorithm relies on two genetic operations, one of them being the mutation process. A single mutation operation is applied to every circuit as follows. First, each individual will mutate, but the block where the mutation happens is randomly chosen. As explained in Section \ref{sec:EG}, this block can be a \textit{non-controlled, controlled} or \textit{helper block}. The mutation mechanism for each type is described below.

\begin{figure}[h]
\begin{center}
\includegraphics[scale=0.3]{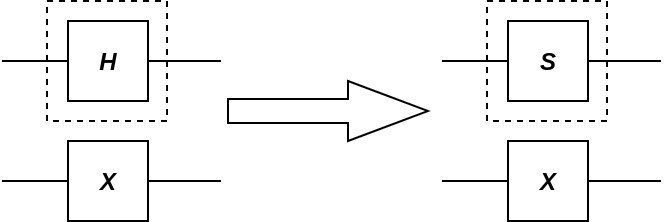}
\caption{Non-controlled block mutation: $H$ mutates to $S$.}
\label{fig:ncmut0}
\end{center}
\end{figure}

For a {\it non-controlled block}, mutation consists of randomly choosing a gate and modifying it by simply replacing it with another randomly chosen gate, as Figure \ref{fig:ncmut0} exemplifies.

\begin{figure}[h]
\begin{center}
\includegraphics[scale=0.28]{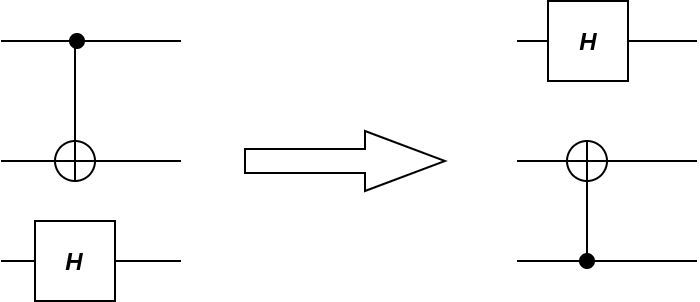}
\caption{Controlled block mutation: random shuffling method.}
\label{fig:cmut0}
\end{center}
\end{figure}

\begin{figure}[h]
\begin{center}
\includegraphics[scale=0.28]{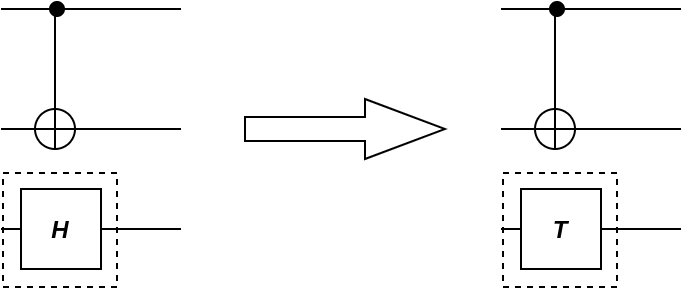}
\caption{Controlled block mutation: gate change method. The $H$ block is changed to a $T$ block.}
\label{fig:cmut1}
\end{center}
\end{figure}

For a \textit{controlled block} three types of mutation are available: \textit{random shuffling of the block},  \textit{removal of a control point}, or \textit{gate change}. The \textit{random shuffling} method just randomly reassigns the control, target, and any other gates present to new positions on the wire. The \textit{removal of a control point} method is only eligible if the number of control points in the block is greater than 1. Since the results discussed here only enforce the use of $X$ as a controlled gate ($CNOT$), this type of mutation has no effect for present purposes but is kept available in case that constraint is removed. The \textit{gate change} method will randomly pick a non-controlled gate (if any) and change its operator to a new randomly chosen gate. If there is no non-controlled gate available, this method is not triggered. If both the shuffling and gate change mutation methods are eligible, one is chosen at random and applied. Figures \ref{fig:cmut0} and \ref{fig:cmut1} show some examples.
\\[12pt]

For a \textit{helper block}, we should note that they are just a sequence of \textit{controlled} and \textit{non-controlled} sub-blocks. For example, the controlled $Z$ block can be viewed as a sequence of a non-controlled $H$ block, a controlled $X$ block, and another non-controlled $H$ block. Given that, in order to mutate a helper block it suffices to randomly pick a sub-block, and mutate it according to its type.

\subsection{Parameters}
\label{section:params}
In all experiments, the first crossover parameters were set to ${\it r_L}=0.7$ and ${\it r_I}=0.3$. The number of population used was $g = 20$, each one with $p = 30$ individuals. The maximum number of iterations allowed was $10^4$. For all experiments, the minimum and maximum block sizes were set to 4 and 15, respectively. 

The quantum gate set available during the search is $\{H,S,CNOT,X,\sqrt{X},T, Z\}$, including also the complex conjugate of the gates $T$, $S$ and $\sqrt{X}$, i.e., the inverse matrices $T^\dagger$, $S^\dagger$ and $\sqrt{X^\dagger}$, respectively. From what we discussed in Section \ref{sec:uniSet} it is clear that the gate set used has more gates needed to provide the universal set for quantum computation. We added extra gates since they are the ones available on different platforms for quantum computation, like the one provided by IBM Quantum Experience~\footnote{http://www.research.ibm.com/quantum/}. Furthermore, they also appear as building blocks in many known algorithms, providing a good trade-off between circuit interpretation, complexity, and search running time
\cite{loceff,lukac1,daskin,lukac1, nielsen_chuang_2010}. 

Gate descriptions and helper blocks used in the search can be found in Appendices \ref{app:Qgates} and \ref{app:helpB}. The matrix representation of the solutions found is given in Appendix \ref{app:sol}.

\section{Results}
\begin{enumerate}
    \item \textbf{Hadamard Coin for the Quantum Walker}
\begin{equation}
\label{eq:HCoin}
U_t = 
\begin{pmatrix} 
1 & 0 & 0 & 0     \\
0 & \frac{1}{\sqrt{2}} & \frac{1}{\sqrt{2}} & 0     \\
0 & \frac{1}{\sqrt{2}} & -\frac{1}{\sqrt{2}} & 0    \\
0 & 0 & 0 & 1
\end{pmatrix}.
\end{equation}

The target unitary above represents the Hadamard coin when it is implemented in tune with the QCA model introduced in \cite{costa2018quantum}. This unitary is also known as the "beam-splitter" operator $F_2$ giving in \cite{cervera2018exact} without the minus factor in the element $\ket{11}\bra{11}$.  Figures \ref{fig:soluc_coin} and \ref{fig:coinplot} show the best circuit found and the evolution curve. In all evolution plots two curves are shown: the blue one shows the trace fidelity of the best leader across all groups (the best individual found so far by the algorithm), and the red one shows the average trace fidelity for all leaders.

The circuit in Figure \ref{fig:soluc_coin} has an error inferior to $1 \times 10^{-16}$ and was found in less than 600 iterations.

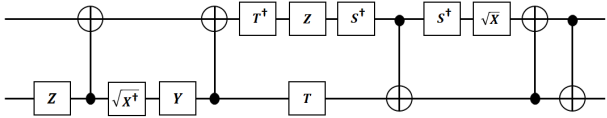
\begin{figure}[htbp]
\centerline{
\Qcircuit @R=1em @C=1.5em {
& \targ  & \gate{S} & \gate{Y} & \targ & \gate{T} & \ctrl{1} & \gate{H} & \ctrl{1} & \gate{Z} & \qw\\
 & \ctrl{-1} & \gate{S} & \gate{H} & \ctrl{-1} & \gate{T} & \targ & \gate{Y}& \targ & \gate{S^\dagger}
& \qw}}
\caption{Solution for the Hadamard coin using the QCA model.\label{fig:soluc_coin}}
\end{figure}

\begin{figure}[h]
\center
\includegraphics[width=20pc]{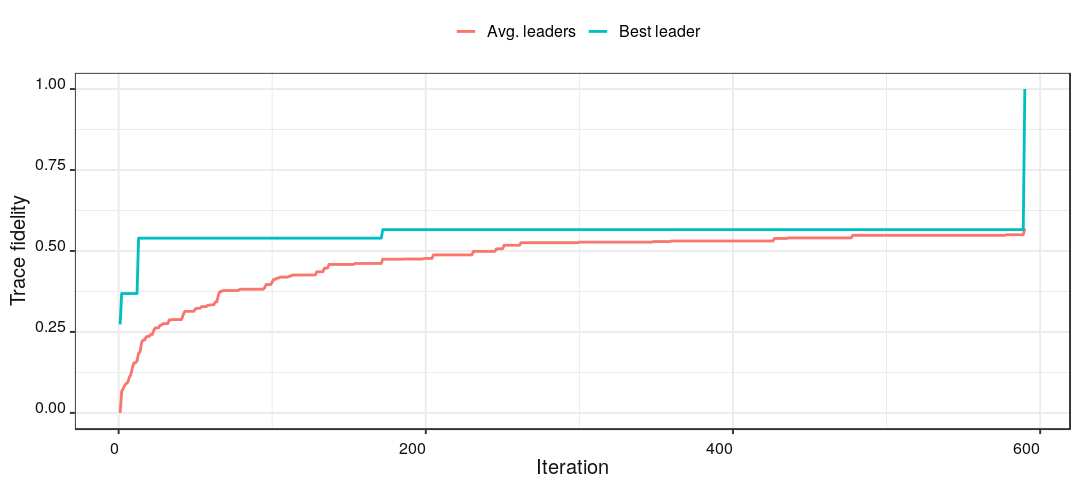}
\caption{Evolution curve for the Hadamard coin.}
\label{fig:coinplot}
\end{figure}

\item \textbf{Toffoli Gate}
\begin{equation}
\label{eq:Tof}
U_t =
\begin{pmatrix} 
1 & 0 & 0 & 0 & 0 & 0 & 0 & 0 \\
0 & 1 & 0 & 0 & 0 & 0 & 0 & 0 \\
0 & 0 & 1 & 0 & 0 & 0 & 0 & 0 \\
0 & 0 & 0 & 1 & 0 & 0 & 0 & 0 \\
0 & 0 & 0 & 0 & 1 & 0 & 0 & 0 \\
0 & 0 & 0 & 0 & 0 & 1 & 0 & 0 \\
0 & 0 & 0 & 0 & 0 & 0 & 0 & 1 \\
0 & 0 & 0 & 0 & 0 & 0 & 1 & 0
\end{pmatrix}.
\end{equation}

The target unitary here is the one in Eq. \eqref{eq:Tof}, namely, the Toffoli gate, which is a controlled-CNOT gate.  Figure \ref{fig:soluc_toff} shows the best circuit found. The solution found has an error inferior to $1 \times 10^{-16}$ and was found in less than 1250 iterations. Figure \ref{fig:toffoliplot} shows the evolution curve during the corresponding optimization process.

\begin{figure}[htbp]
\centerline{
\Qcircuit @R=2em @C=0.4em {
& \qw & \ctrl{2} & \qw& \ctrl{2} & \gate{T^{\dagger}} & \targ & \gate{T} & \ctrl{2}& \qw&\ctrl{2} & \qw & \qw & \qw & \qw & \targ & \qw \\
 & \qw & \qw & \qw & \qw & \qw  & \ctrl{-1} & \qw& \qw & \qw& \qw &\ctrl{1} & \qw & \ctrl{1} & \gate{T^{\dagger}} & \ctrl{-1} & \qw \\
 & \gate{H} & \targ & \gate{T} & \targ & \qw & \qw & \qw&\targ & \gate{T^\dagger}& \targ & \targ & \gate{T} & \targ & \gate{T^\dagger} & \gate{H} & \qw 
}}
\caption{Solution for the Toffoli unitary.\label{fig:soluc_toff}}
\end{figure}
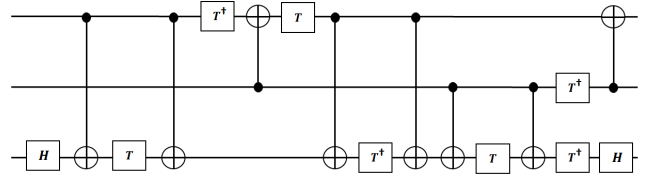

\begin{figure}[h]
\center
\includegraphics[width=20pc]{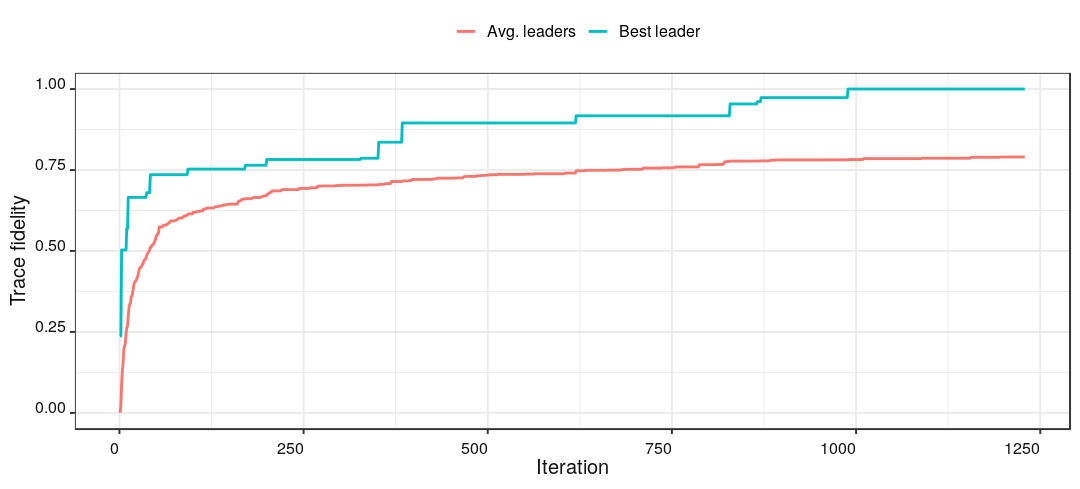}
\caption{Evolution curve for the Toffoli unitary.}
\label{fig:toffoliplot}
\end{figure}

\item \textbf{Fredkin Gate}
\begin{equation}
U_t =
\begin{pmatrix} 
1 & 0 & 0 & 0 & 0 & 0 & 0 & 0 \\
0 & 1 & 0 & 0 & 0 & 0 & 0 & 0 \\
0 & 0 & 1 & 0 & 0 & 0 & 0 & 0 \\
0 & 0 & 0 & 1 & 0 & 0 & 0 & 0 \\
0 & 0 & 0 & 0 & 1 & 0 & 0 & 0 \\
0 & 0 & 0 & 0 & 0 & 0 & 1 & 0 \\
0 & 0 & 0 & 0 & 0 & 1 & 0 & 0 \\
0 & 0 & 0 & 0 & 0 & 0 & 0 & 1
\end{pmatrix}.
\end{equation}
The unitary above represents the Fredkin gate, the subsequent target unitary and Figure \ref{fig:soluc_fred} shows the best circuit found. The solution found has an error inferior to $1 \times 10^{-16}$ and was found in less than 5000 iterations. Figure \ref{fig:fredkinplot} shows the best circuit found and the evolution curve.

\begin{figure}[htbp]
\centerline{
\Qcircuit @R=2em @C=0.3em {
& \qw &\qw &\targ & \gate{T}&\ctrl{2} \qw & \qw & \ctrl{2} & \targ &\qw &\qw &\ctrl{2} & \qw & \ctrl{2}& \gate{T^\dagger} &\qw & \qw & \qw &\qw &\qw &\qw\\
& \qw & \targ & \ctrl{-1} & \qw & \qw & \qw &\qw &\ctrl{-1} &\ctrl{1}& \qw & \qw & \qw & \qw & \gate{T} & \ctrl{1}&\qw & \ctrl{1} &\qw & \targ &\qw\\
&\qw & \ctrl{-1} \qw & \gate{H}& \gate{T}&\targ &\gate{T^{\dagger}} & \targ & \gate{H} & \targ & \gate{H}& \targ & \gate{T} & \targ & \qw & \targ & \gate{T^\dagger} & \targ & \gate{H} & \ctrl{-1}&\qw
}}
\caption{Solution for the Fredkin unitary.\label{fig:soluc_fred}}
\end{figure}
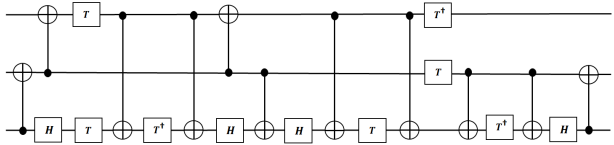

\begin{figure}[h]
\center
\includegraphics[width=20pc]{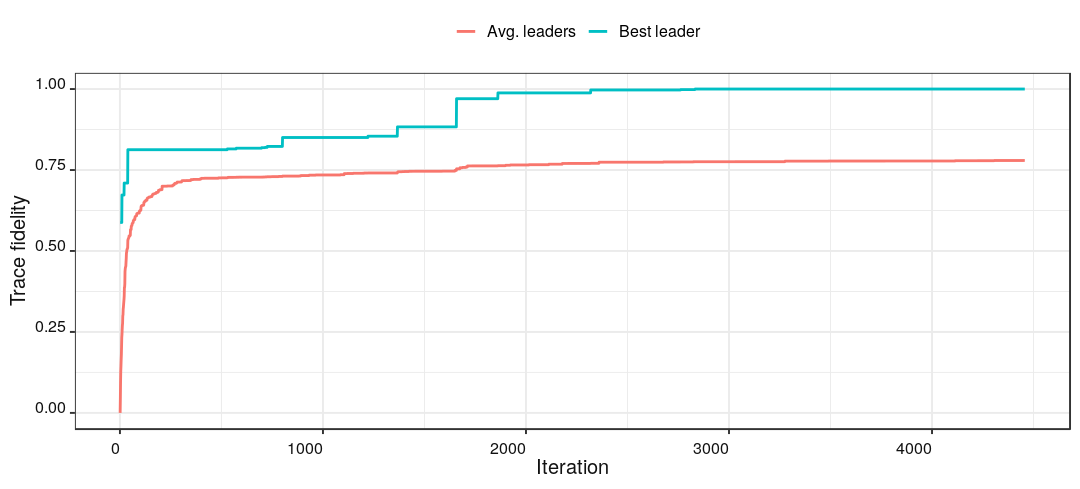}
\caption{Evolution curve for the Fredkin unitary.}
\label{fig:fredkinplot}
\end{figure}
\end{enumerate}
\section{Concluding Remarks}
The automatic synthesis of quantum circuits is an open and important problem for the development of quantum algorithms. This work explores the use of an evolutionary search algorithm that uses groups of populations that evolve separately, subsequently exchanging genetic material with each other. In our experiments, classical genetic algorithms seems to suffer from the problem of premature convergence, with the members of the population becoming very similar or equal, leading to a local minimum in most cases. 

Comparing our results for the Toffoli gate decomposition with its most cited implementation (\cite{nielsen_chuang_2010}, page 182), we obtained the same amount of non-Clifford gates, which are the most expensive gates for the fault-tolerant quantum computers. However, we could not do better than \cite{jones2013low}, in terms of non-Clifford gates, an implementation that requires ancillary qubits. Now, contrasting the results established here with the Fredkin gate, which is a controlled-SWAP gate, we obtained the same number of non-Clifford gates used in the Toffoli decomposition. Our results, then, agree with the decomposition discussed in \cite{nielsen_chuang_2010}, which shows how to implement Fredkin gate with just one single Toffoli gate, rather than three. 

A comparison for the case of the Hadamard we have the results from \cite{cervera2018exact}. In their circuit implementation equivalent to  equation \ref{eq:HCoin} they have the same depth and the same number of non-Clifford gates of . On the other hand we have a higher number of total gates which includes one extra CNOT.

The results showed here are remarkable in the sense that we proposed a non specialized approach for the decomposition method for controlled gates \cite{barenco1995elementary} and we did not apply any specific theory of circuit decomposition, except the concept of universal gates. Furthermore, our implementation was able to find the solutions in a number of iterations that were either smaller or comparable to other analogous works \cite{daskin,lukac1}.

One important limitation of the present work relates to the maximum number of qubits allowed: unitaries with 10 or more qubits will make the search considerably slower, due to the exponential usage of memory. But this can be relatively easy to remedy by making use of sparse data structures.

As expected, the size of the unitary seems to affect the number of iterations until a solution is found (all the search parameters were hold constant across the experiments). The solutions for the Toffoli and Fredkin unitaries took 1250 and 5000 iterations, respectively, against 600 for the smaller quantum walker coin unitary. Despite having the same dimension, the Fredkin unitary seems more challenging to the search algorithm, requiring four times the number of iterations, which is expected since its ordinary decomposition has higher depth than the Toffoli gate~\cite{smolin1996five}. Also, as we pointed out before, the solutions presented redundant circuit segments, e.g., a wire with two Hadamard gates in sequence, resulting in $H H = I$. In a subsequent undertaking, we plan to implement a post-processing stage to the whole process in order to replace the visual inspection we performed for the results shown.

In contrast with results obtained in similar works, two aspects stand out positively in this work: the generic character of the approach described and the flexibility of the solutions found. The presented algorithm is readily applicable to any target unitary, with added flexibility for specifying the set of gates and helper blocks available for the search, and also, the minimum and maximum sizes for the possible solutions.

It would also be fruitful to pursue further research about the efficiency of the genetic operators presented and investigate whether and how the search space can be decreased in an efficient way. This could help the search for optimal circuits, which are important when taking into account the cost of the physical implementation of these components. For instance, during the circuit decomposition process, it is feasible to look for a gate set that decreases the time consumption of a quantum computation on fault-tolerant quantum computers.   We can also imagine a scenario of applying this method for circuit synthesis that uses more complex transformations, e.g.,  Toffoli and Fredkin gates, or even Fourier transformation, rather than the basic gate set employed herein. In this scenario, our approach would help us decompose more complicated unitaries applied to practical problems (like quantum algorithms for differential equations \cite{costa2019quantum,berry2014high} or combinatorial optimization problems~\cite{PRXQuantum.1.020312}); this would also provide good estimation of necessary non-Clifford gates for those applications, which are the most complicated gates for the fault-tolerant quantum devices.

Finally, we can point out that a faster convergence can be achieved by introducing more high-level components into the search space.  We believe that the built-in flexibility of our approach (without loss of performance), together with the scalability factor provided by the island model, provides valuable insight into the study of automatic quantum circuit synthesis for the currently available quantum technologies. Therefore, this work provided the proof of principle that the island GA can be used for circuit synthesis, regardless of the practical quantum device used.

\section*{Acknowledgements}
This work was made possible thanks to grants provided by CAPES (Mackenzie PriInt project 88887.310281/2018-00 and STIC-AmSud CoDANet project 88881.197456/2018-01) and by CNPq-PQ 305199/2019-6. P.C.S.C. wishes to thank Gavin Brennen for his helpful comments.

\section{Solutions Found}
\label{sec:solF}
\label{app:sol}
\begin{enumerate}
\item[]{\it Hadamard Coin via QCA} \\
\begin{equation}
U_a=
\begin{pmatrix} 
1 & 3.925 \times 10^{-17} & 3.925 \times 10^{-17} & 0 \\
5.55111 \times 10^{-17} & 7.071 \times 10^{-1} & 7.071 \times 10^{-1} & 0 \\
0 & 7.071 \times 10^{-1} & -7.071 \times 10^{-1} & 0 \\
0 & 0 & 0 & 1
\end{pmatrix}. 
\end{equation}

\item[]{\it Toffoli Gate}
\begin{equation}
U_a=
\begin{pmatrix} 
1 & 0 & 0 & 0 & 0 & 0 & 0 & 0 \\
0 & 1 & 0 & 0 & 0 & 0 & 0 & 0 \\
0 & 0 & 1 & 0 & 0 & 0 & 0 & 0 \\
0 & 0 & 0 & 1 & 0 & 0 & 0 & 0 \\
0 & 0 & 0 & 0 & 1 & -1.11 \times 10^{-16}\mathrm{i} & 0 & 0 \\
0 & -1.11 \times 10^{-16}\mathrm{i} & 0 & 0 & 0 & 1 & 0 & 0 \\
0 & 0 & 0 & 0 & 0 & 0 & 0 & 1 \\
0 & 0 & 0 & 0 & 0 & 0 & 1 & 0
\end{pmatrix} .
\end{equation}

\item[]{\it Fredkin Gate}
\begin{equation}
U_a=
\begin{pmatrix} 
1 & 0 & 2.36 \times 10^{-16} & 0 & 0 & 0 & 0 & 0 \\
0 & 1 & 0 & 0 & 0 & 0 & 0 & 0 \\
0 & 0 & 1 & 0 & 0 & 0 & 0 & 0 \\
2.36 \times 10^{-16} & 0 & 0 & 1 & 0 & 0 & 0 & 0 \\
0 & 0 & 0 & 0 & 1 & 0 & 0 & 0 \\
0 & 0 & 0 & 0 & 0 & 0 & 1 & 0 \\
0 & 0 & 0 & 0 & 0 & 1 & 0 & 0 \\
0 & 0 & 0 & 0 & 0 & 0 & 0 & 1
\end{pmatrix}. 
\end{equation}

\section{Quantum gates}
\label{app:Qgates}
The following is a list of the gates employed in the work. All rotations described below are measured in radians.

\item[] {\it Hadamard gate} \\
Performs a $\pi$ rotation about the $Z$-axis, followed by a $\frac{\pi}{2}$ rotation about the Y-axis.
\begin{equation}
H = \frac{1}{\sqrt{2}} \begin{pmatrix}  1 & 1 \\ 1 & -1 \end{pmatrix}.
\end{equation}

\item[] {\it Pauli-$X$ gate} \\
Performs a $\pi$ rotation about the X-axis.
\begin{equation}
X = \begin{pmatrix} 0 & 1 \\1 & 0 
\end{pmatrix}.
\end{equation}

\item[] {\it Pauli-$Y$ gate} \\
Performs a $\pi$ rotation about the $Y$-axis.
\begin{equation}
Y = \begin{pmatrix} 0 & -\mathrm{i} \\\mathrm{i} & 0 
\end{pmatrix}.
\end{equation}

\item[] {\it Pauli $Z$-gate} \\
Performs a $\pi$ rotation about the $Z$-axis.
\begin{equation}
Z = \begin{pmatrix} 1 & 0 \\0 & -1
\end{pmatrix}.
\end{equation}

\item[] {\it $S$ gate} \\
The phase gate, performs a $\frac{\pi}{2}$ rotation about the $Z$-axis.
\begin{equation}
S = \begin{pmatrix} 1 & 0 \\0 & \mathrm{i}
\end{pmatrix}.
\end{equation}

\item[] {\it $T$ gate} \\
Performs a $\frac{\pi}{8}$ rotation about the $Z$-axis.
\begin{equation}
T =  \begin{pmatrix}  1 & 0 \\ 0 & \frac{1+ \mathrm{i}}{\sqrt{2}}\end{pmatrix}
\end{equation}.

\item[] {\it $\sqrt{X}$ gate} \\
Square root of the $X$-Pauli gate.

\begin{equation}
\sqrt{X} = \frac{1}{2} \begin{pmatrix}  1 + \mathrm{i} & 1 -  \mathrm{i} \\ 1 - \mathrm{i} & 1 + \mathrm{i} \end{pmatrix}.
\end{equation}
\end{enumerate}

\section{Helper blocks}
\label{app:helpB}

The following is the list of helper blocks the algorithm described can rely on.

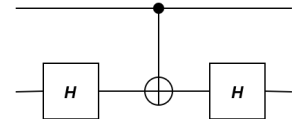
\begin{figure}[!h]
\centerline{
\Qcircuit @R=2em @C=1em {
& \qw &\ctrl{1} & \qw & \qw \\
 & \gate{H} & \targ & \gate{H} & \qw 
}}
\caption{Controlled $Z$ block.}
\end{figure}

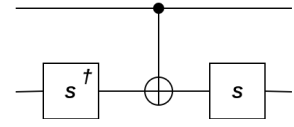
\begin{figure}[!h]
\centerline{
\Qcircuit @R=2em @C=1em {
& \qw &\ctrl{1} & \qw & \qw \\
 & \gate{S^{\dagger}} & \targ & \gate{S} & \qw 
}}
\caption{Controlled $Y$ block.}
\end{figure}

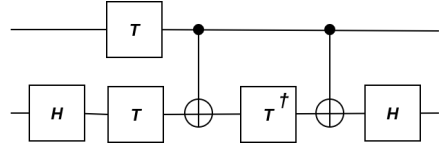
\begin{figure}[!h]
\centerline{
\Qcircuit @R=2em @C=1em {
& \qw & \gate{T} &\ctrl{1} & \qw &\ctrl{1} & \qw &\qw\\
 & \gate{H} & \gate{T} &\targ & \gate{T^{\dagger}}  &\targ & \gate{H} &\qw
}}
\caption{Controlled $\sqrt {X}$ block.}
\end{figure}

\begin{figure}[!h]
\centerline{
\Qcircuit @R=2em @C=1em {
& \qw  &\ctrl{1} & \gate{T} &\qw &\ctrl{1} & \qw \\
 & \gate{T}  &\targ & \gate{S^{\dagger}}  & \gate{T}&\targ & \qw
}}
\caption{Controlled $S$ block.}
\end{figure}
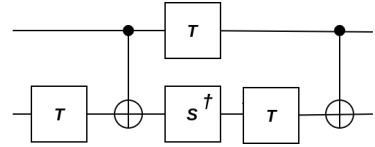

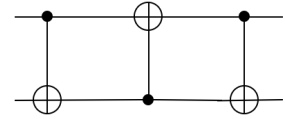
\begin{figure}[!h]
\centerline{
\Qcircuit @R=2em @C=1em {
& \ctrl{1} &\targ & \ctrl{1} & \qw \\
 & \targ & \ctrl{-1} & \targ & \qw 
}}
\caption{Swap block.}
\end{figure}

\end{document}